\documentclass[journal]{IEEEtran}

\usepackage[OT1]{fontenc} 
\usepackage[numbers,sort&compress]{natbib}
\usepackage[cmex10]{amsmath}
\usepackage{amssymb}
\usepackage{bm}
\usepackage[dvipdfmx]{graphicx}
\usepackage{color}
\usepackage{subfigure}
\usepackage{tabularx}
\usepackage{arydshln}
\usepackage{mathtools}
\usepackage{multirow}
\usepackage{algorithm,algorithmic}
\usepackage{url}
\usepackage{listings}
\usepackage{comment}
\usepackage{latexsym}
\usepackage{amsthm}
\usepackage{braket}
\usepackage{physics}
\usepackage{mleftright}
\mleftright
\usepackage{hyperref}
\usepackage[absolute,overlay]{textpos}

\def\y{\mathbf{y}}
\def\H{\mathbf{H}}

\def\x{\mathbf{x}}
\def\s{\mathbf{s}}
\def\n{\mathbf{n}}
\def\A{\mathbf{A}}
\def\R{\mathbb{R}}
\def\I{\mathbb{I}}

\def\E{\mathbb{E}}

\def\j{\mathrm{j}}

\makeatletter
\def\ps@IEEEtitlepagestyle{%
  \def\@oddfoot{\mycopyrightnotice}%
  \def\@oddhead{\hbox{}\@IEEEheaderstyle\leftmark\hfil\thepage}\relax
  \def\@evenhead{\@IEEEheaderstyle\thepage\hfil\leftmark\hbox{}}\relax
  \def\@evenfoot{}%
}
\def\mycopyrightnotice{%
  \begin{minipage}{\textwidth}
  \centering \scriptsize
\textcopyright 2023 IEEE. Personal use of this material is permitted.
  Permission from IEEE must be obtained for all other uses, in any current or future
  media, including reprinting/republishing this material for advertising or promotional
  purposes, creating new collective works, for resale or redistribution to servers or
  lists, or reuse of any copyrighted component of this work in other works.
  DOI: \href{https://doi.org/10.1109/LWC.2023.3310586}{10.1109/LWC.2023.3310586}
  \end{minipage}
}
\makeatother
\begin{document}
\title{\huge On the Capacity of Generalized Quadrature Spatial Modulation}

\author{Kein~Yukiyoshi and Naoki~Ishikawa,~\IEEEmembership{Senior~Member,~IEEE}.\thanks{K.~Yukiyoshi and N.~Ishikawa are with the Faculty of Engineering, Yokohama National University, 240-8501 Kanagawa, Japan (e-mail: ishikawa-naoki-fr@ynu.ac.jp). This work was partially supported by Support Center for Advanced Telecommunications Technology Research, Japan.}}

\maketitle

\TPshowboxesfalse
\begin{textblock*}{\textwidth}(45pt,10pt)
\footnotesize
\centering
Accepted for publication in IEEE Wireless Communications Letters. This is the author's version which has not been fully edited and content may change prior to final publication. Citation information: DOI 10.1109/LWC.2023.3310586
\end{textblock*}

\begin{abstract}
In this letter, the average mutual information (AMI) of generalized quadrature spatial modulation (GQSM) is first derived for continuous-input continuous-output channels. Our mathematical analysis shows that the calculation error induced by Monte Carlo integration increases exponentially with the signal-to-noise ratio. This nature of GQSM is resolved by deriving a closed-form expression. The derived AMI is compared with other related SM schemes and evaluated for different antenna activation patterns. Our results show that an equiprobable antenna selection method slightly decreases AMI of symbols, while the method significantly improves AMI in total.
\end{abstract}

\begin{IEEEkeywords}
Capacity, mutual information, spatial modulation (SM), quadrature spatial modulation (QSM), generalized quadrature spatial modulation (GQSM).
\end{IEEEkeywords}

\IEEEpeerreviewmaketitle

\section{Introduction}
Spatial modulation (SM) is a technique that modulates information by assigning it to an index of active transmit antennas, in addition to data symbols \cite{ishikawa201850}. SM has been extensively studied as a potential solution for striking the fundamental trade-off between performance and complexity in wireless communications \cite{wen2019survey}.

The transmission rate of SM is given by $R_{\mathrm{SM}} = \log_2 L + \lfloor\log_2 N_t\rfloor$, where $L$ is the constellation size and $N_t$ is the number of transmit antennas. To improve the spectral efficiency of SM, a number of extensions have been proposed.
Introducing representative schemes, generalized spatial modulation (GSM)~\cite{jeganathan2008generalized} extends the number of data symbols from $1$ to a general integer $K$, resulting in an improved transmission rate $R_{\mathrm{GSM}} = K\log_2 L + \lfloor\log_2 \binom{N_t}{K}\rfloor$. In contrast, quadrature spatial modulation (QSM)~\cite{mesleh2015quadrature} defines different activation patterns (APs) independently for the real and imaginary parts of the codeword, resulting in an improved transmission rate $R_{\mathrm{QSM}} = \log_2 L + 2 \lfloor\log_2 N_t\rfloor$. A hybrid of the above two SM extensions,  generalized quadrature spatial modulation (GQSM)~\cite{castillo-soria2017generalized}, has been proposed. Currently, GQSM is considered to be the most advanced SM, offering the highest transmission rate $R_{\mathrm{GQSM}} = K\log_2L + 2 \lfloor\log_2\binom{N_t}{K}\rfloor$.
Additionally, other equivalent techniques have been proposed in the context of orthogonal frequency division multiplexing (OFDM), generally termed index modulation (IM), such as OFDM-IM \cite{abu-alhiga2009subcarrierindex,basar2013orthogonal} and OFDM-I/Q-IM \cite{zheng2015lowcomplexity}.

GQSM requires APs of antennas to be designed carefully, where $Q$ APs are selected out of $\binom{N_t}{K}^2$ possible candidates. This AP selection determines achievable performances, leading to studies on the efficient design of APs~\cite{basar2013orthogonal,wen2016equiprobable,ishikawa2019imtoolkit}. One approach, known as combinatorial design, was proposed in \cite{basar2013orthogonal}; it is equivalent to selecting $Q$ APs from $\binom{N_t}{K}^2$ candidates in lexicographic order. Another approach, known as equiprobable design, was proposed in \cite{wen2016equiprobable}; APs are constructed such that each transmit antenna is activated with an equal probability. In addition, an integer linear programming (ILP) design is proposed in \cite{ishikawa2019imtoolkit}, where the equiprobable antenna selection is formulated as an ILP problem and it is compared with other design methods \cite{basar2013orthogonal,wen2016equiprobable} for discrete-input channels.\footnote{The open-source implementations of \cite{basar2013orthogonal,wen2016equiprobable,ishikawa2019imtoolkit} are provided in \cite{ishikawa2019imtoolkit}.}

The channel capacity $C$ is an essential metric for evaluating a communication system, and many studies have addressed that analysis of SM.
Although the distribution of the channel input that maximizes the average mutual information (AMI) of SM is unknown~\cite{basnayaka2016massive}, the AMI of SM and GSM has been derived under the assumption of Gaussian input distribution~\cite{shamasundar2022capacity}.
Under the same assumption, it was shown in \cite{younis2017quadrature} that, when $N_t \rightarrow \infty$, the AMI of QSM is equal to the channel capacity of MIMO~\cite{telatar1999capacity}.
To the best of our knowledge, the AMI of QSM or GQSM at a specific number of transmit antennas has yet to be derived, although it is an important metric predicting an upper bound of achievable rates in coded scenarios.

In this letter, we newly derive the AMI of GQSM assuming continuous-input channels and compare it with those of SM and GSM, where a non-trivial problem of calculation errors is solved by our partially closed-form expressions. In addition, using the derived AMI, we investigate differences between the three methods \cite{basar2013orthogonal,wen2016equiprobable,ishikawa2019imtoolkit} of APs and clarify that the difference is maximized at medium signal-to-noise ratios (SNRs).

\section{System Model}
Consider an $N_t\times N_r$ multiple-input multiple-output (MIMO) system and assume independent and identically distributed
(i.i.d.) frequency-flat Rayleigh fading channels in which each element of a channel matrix $\mathbf{H} \in \mathbb{C}^{N_r \times N_t}$ and a noise vector $\mathbf{n} \in \mathbb{C}^{N_r \times 1}$ independently follow complex Gaussian distributions $\mathcal{CN}(0, 1)$ and $\mathcal{CN}(0, \sigma_n^2)$, respectively.
Given the noise variance $\sigma_n^2$ and the transmission power $\sigma_s^2$, the SNR is defined by $\rho = \sigma_s^2/\sigma_{n}^2$.
Let $x_i$ be the $i$-th element of a GQSM codeword $\x \in \mathbb{C}^{N_t \times 1}$.
The real and imaginary parts of classic symbols $\s \in \mathbb{C}^{K \times 1}$ are denoted by $\s_{\R}$ and $\s_{\I}$, while the $k$-th elements of $\s_{\R}$ and $\s_{\I}$ are denoted by $s_{\R}^{(k)}$ and $s_{\I}^{(k)}$, respectively.
Similarly, the $(i, k)$ element of a matrix $\A$ is denoted by $a^{(i, k)}$.
Then, the APs corresponding to the real and imaginary parts of the codeword are defined by
\begin{align}
    \mathbb{A}_\R = \bigl\{ \A_\R \in \qty{0, 1}^{N_t \times K} \mid \forall k = 1, \cdots K,\ \Sigma_{i=1}^{N_t} a_{\R}^{(i, k)} = 1 \bigr\}
\end{align}
and
\begin{align}
    \mathbb{A}_\I = \bigl\{\A_\I \in \qty{0, 1}^{N_t \times K} \mid \forall k = 1, \cdots K,\ \Sigma_{i=1}^{N_t} a_{\I}^{(i, k)} = 1\bigr\},
\end{align}
where we have relationships $a_{\R}^{(i, k)} = 1 \Leftrightarrow \Re(x_{i}) = s_{\R}^{(k)}$ and $a_{\I}^{(i, k)} = 1 \Leftrightarrow \Im(x_{i}) = s_{\I}^{(k)}$. Denoting all APs as $\mathbb{A} = \mathbb{A}_\R \times \mathbb{A}_\I$ using the Cartesian product $\times$, the number of APs, $Q = |\mathbb{A}|$, satisfies a constraint
$2 \leq Q \leq 2^{\lfloor\log_2\binom{N_t}{K}^2\rfloor} \leq \binom{N_t}{K}^2$
for additional bit allocation of GQSM.\footnote{Here, $Q$ is not limited to the maximum value, which can be adjusted to achieve additional gain at the expense of a reduced transmission rate \cite{ishikawa2019imtoolkit}.}
The received signal $\y$ is represented as
\begin{equation}
    \y = \H\x + \mathbf{n} \in \mathbb{C}^{N_r \times 1}
\end{equation}
and the codeword $\x$ is constructed by
\begin{equation}
    \x = \A_\R\s_\R + \j\A_\I\s_\I \in \mathbb{C}^{N_t \times 1},
\end{equation}
where $\j = \sqrt{-1}$ denotes the imaginary number. Note that this generalized system model can represent QSM by imposing a constraint $K=1$ and GSM by imposing a condition $\A_\R = \A_\I$ for all APs $(\A_\R, \A_\I) \in \mathbb{A}$. In addition, by setting the off-diagonal elements of the channel matrix $\H$ to $0$, it becomes equivalent to an idealized system model of OFDM-I/Q-IM.

\section{Capacity Analysis}
AMI $I(\x ; \y | \H)$ is defined as the expected value of the maximum number of bits that can be conveyed without error at a given SNR, and ergodic capacity $C = \max_{p(\x)} I(\x ; \y | \H)$ is defined as the maximum value of AMI over the distribution of the codeword $\x$~\cite{telatar1999capacity}.
In general, the ergodic capacity of MIMO is achieved when each element of the codeword $\x$ independently follows a complex Gaussian distribution with the same variance~\cite{telatar1999capacity}.
However, a distribution that achieves the capacity of SM has not yet been found~\cite{basnayaka2016massive}. Therefore, in this study, we derive the AMI of GQSM when the input symbols independently follow a complex Gaussian distribution with the same variance as in previous studies~\cite{shamasundar2022capacity} and use it as an evaluation metric instead of the actual capacity.

\subsection{AMI of Discrete-Input Channel~\cite{ng2006mimo}}
If the input symbols follow a discrete probability distribution, the AMI of a general MIMO scheme is expressed as~\cite{ng2006mimo}
\begin{equation}
    I(\x ; \y|\H) = R - \frac{1}{2^R}\sum_{i=1}^{2^R}\E_{\H,\n}\qty[\log_2\sum_{j=1}^{2^R}\exp\eta[i, j]], \label{eq:I_discrete}
\end{equation}
where
\begin{equation}
    \eta[i, j] = \frac{-\norm{\H(\x_i - \x_j) + \n}^2 + \norm{\n}^2}{\sigma_n^2},
\end{equation}
$\x_i$ is the $i$-th element of a codebook $\qty{\x_1 \cdots, \x_{2^R}}$, and $R$ is the transmission rate.
\subsection{AMI of Continuous-Input Channel}
In the following, the AMI of GQSM is newly derived for continuous-input continuous-output channels. If the input symbols follow a continuous probability distribution, similar to \cite[Eq.~(31)]{shamasundar2022capacity}, the AMI of GQSM can be divided into that of the symbols, $I_\s$, and that of the APs, $I_\A$, expressed as
\begin{align}
    \label{eq:separate}
    I(\A, \s ; \y|\H) = I(\s ; \y | \A, \H) + I(\A ; \y | \H) = I_\s + I_\A.
\end{align}
First, we derive $I_\s$. From \eqref{eq:separate}, we obtain
\begin{align}
    \label{eq:I_s}
  I_\s  &= I(\s ; \y | \A, \H) = h(\y|\A, \H) - h(\y|\A, \H, \s) \nonumber\\
  &= -\E_{\A, \H, \y}\qty[\log_2 p(\y | \A, \H)] - h(\n) \nonumber\\
  &= \E_{\A, \H, \y}\qty[\log_2 p(\y | \A, \H)^{-1}] - N_r\log_2(\pi e \sigma_n^2),
\end{align}
where $h(\cdot)$ denotes the entropy of a random variable. Supposing $d\s = ds_{\R}^{(1)} \cdots ds_{\R}^{(k)} ds_{\I}^{(1)} \cdots ds_{\I}^{(k)}$, the argument of the expectation $\E_{\A, \H, \y}[\cdot]$ can be transformed into a Monte Carlo integration~\cite{ishikawa201850,basnayaka2016massive} of
\begin{align}
    \label{eq:p_y_AH1}
    &\log_2 p(\y|\A, \H)^{-1} = -\log_2 \int_{\mathbb{R}^{2K}} p(\s)p(\n = \y - \H\x)\ d\s \nonumber\\
    =& -\log_2 \qty(\E_\s\qty[p(\n = \y - \H\x)]) \nonumber\\
    =& -\log_2\qty(\E_\s\qty[\exp\qty(- \frac{\norm{\y - \H\x}^2}{\sigma_n^2})]) + N_r\log_2\qty(\pi\sigma_n^2).
\end{align}
Substituting \eqref{eq:p_y_AH1} for \eqref{eq:I_s} yields
\begin{equation}
    \label{eq:I_s2}
    I_\s = \E_{\A, \H, \y}\qty[-\log_2 \E_\s\qty[\exp\qty(- \frac{\norm{\y - \H\x}^2}{\sigma_n^2})]] - N_r\log_2(e).
\end{equation}
Second, we derive $I_\A$. From \eqref{eq:separate}, we obtain
\begin{align}
    I_\A &= I(\A ; \y | \H) = h(\A | \H) - h(\A | \H, \y) \nonumber\\
    &= \log_2Q - \E_{\A, \H, \y}\qty[\log_2p(\A | \y, \H)^{-1}]. \label{eq:I_A}
\end{align}
Using Bayes' theorem, $p(\A | \y, \H)$ is expressed as
\begin{align}
    p(\A | \y, \H) &= \frac{p(\y | \A, \H)p(\A | \H)}{\sum_{i=1}^Qp(\y | \A_i, \H)p(\A_i | \H)}\nonumber\\
    &= \frac{p(\y | \A, \H)p(\A)}{\sum_{i=1}^Qp(\y | \A_i, \H)p(\A_i)}.
\end{align}
Assuming that APs are chosen uniformly at random, i.e., $p(\A_i) = 1/Q$, we obtain
\begin{align}
    p(\A | \y, \H) 
    &= \frac{p(\y | \A, \H)}{\sum_{i=1}^Qp(\y | \A_i, \H)} \nonumber\\
    &= \frac{\E_\s\qty[\exp\qty(-\frac{\norm{\y - \H\x}^2}{\sigma_n^2})]}{\sum_{i=1}^Q \E_\s\qty[\exp\qty(- \frac{\norm{\y - \H\x_i}^2}{\sigma_n^2})]}.
    \label{eq:p_l_yH}
\end{align}
Substituting \eqref{eq:p_l_yH} for \eqref{eq:I_A} yields
\begin{equation}
\label{eq:I_A2}
\begin{split}
        I_\A &= \log_2Q\\
        &\quad - \E_{\A, \H, \y}\qty[\log_2 \frac{\sum_{i=1}^Q \E_\s\qty[\exp\qty(- \frac{\norm{\y - \H\x_i}^2}{\sigma_n^2})]}{\E_\s\qty[\exp\qty(-\frac{\norm{\y - \H\x}^2}{\sigma_n^2})]}].
\end{split}
\end{equation}
Overall, from \eqref{eq:I_s2} and \eqref{eq:I_A2}, the AMI of GQSM for continuous-input channels is derived as
\begin{align}
\label{eq:I_continuous}
&I(\x ; \y|\H) = I_\s + I_\A \nonumber\\
     &= -\E_{\A, \H, \y}\qty[\log_2 \sum_{i=1}^Qp(\y | \A_i, \H)] + \log_2Q - N_r\log_2\qty(e) \nonumber \\
     &= -\E_{\A, \H, \y}\qty[\log_2\sum_{i=1}^Q \E_\s\qty[\exp\qty(-\frac{\norm{\y - \H\x_{i}}^2}{\sigma_n^2})]] \nonumber \\
     &\quad + \log_2Q - N_r\log_2\qty(e),
\end{align}
where the Monte Carlo method is used to calculate the expected values $\E_{\A, \H, \y}[\cdot]$ and $\E_\s[\cdot]$.

\subsection{Error Analysis of Monte Carlo Integration}
\begin{figure}[tb]
    \centering
\includegraphics[keepaspectratio,scale=0.69]{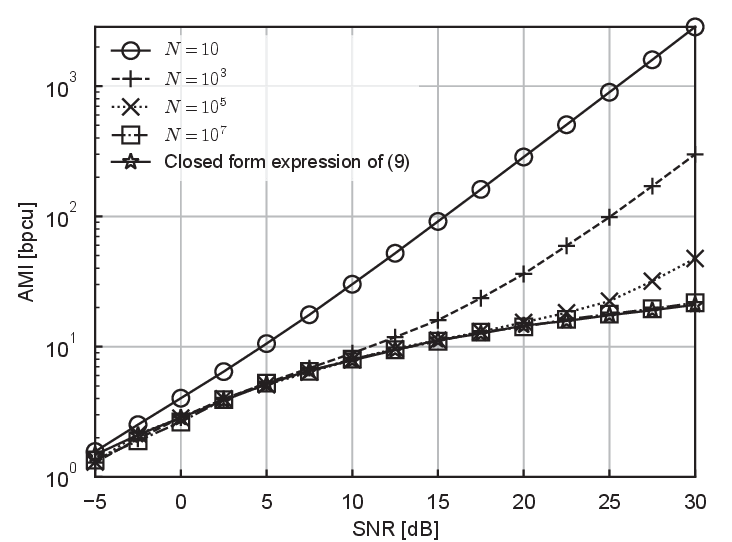}
    \caption{Relationship between SNR and $I_\s$ calculated from \eqref{eq:I_s2}.}
    \label{fig:AMI_error}
\end{figure}
The expression in \eqref{eq:I_continuous} seems straightforward; however, it is actually difficult to calculate with high accuracy. Fig.~\ref{fig:AMI_error} shows $I_\s$ calculated from \eqref{eq:I_s2} using the Monte Carlo integration for different sample sizes $N=10, \cdots, 10^7$, where $(N_t, N_r, K, \mathbf{A}_\R, \mathbf{A}_\I) = (2, 2, 1, [0\ 1]^\mathrm{T}, [1\ 0]^\mathrm{T})$, and each element of the input symbol follows a complex Gaussian distribution $\mathcal{CN}(0, \sigma_s^2/2K)$ independently. 
$I_s$ calculated using the closed form expression of \eqref{eq:p_y_AH1} is also shown, which will be discussed in Section~\ref{subsec:closed-form}.
As shown in Fig.~\ref{fig:AMI_error}, the approximate value of $I_\s$ increases exponentially from a certain SNR if the sample size is relatively small.
This result indicates that the calculation errors increase exponentially with respect to SNR, which is analyzed below.

First, the expectation of \eqref{eq:p_y_AH1} is transformed into another form of
\begin{align}
    \label{eq:p_y_AH_numerical}
    &\log_2 \bar{p}(\y | \A, \H)^{-1} \nonumber\\
    =& -\log_2\qty(\frac{1}{N}\sum_{i=1}^N\frac{1}{\qty(\pi \sigma_n^2)^{N_r}}\exp\qty(- \frac{\norm{\y - \H\x^{(i)}}^2}{\sigma_n^2})) \nonumber\\
    =& -\log_2\qty(\sum_{i=1}^N\exp\qty(- \frac{\norm{\y - \H\x^{(i)}}^2}{\sigma_n^2})) \nonumber \\
    &+ \log_2(N\pi^{N_r}\sigma^{2N_r}).
\end{align}
Here, the variables $\y$, $\A$, and $\H$ in \eqref{eq:p_y_AH1} other than $\s$ can be regarded as given constants. Therefore, because of the reproductive property of the Gaussian distribution, if each element of the input symbol follows a complex Gaussian distribution independently, then $\y - \H\x$ is also a probability vector whose elements follow complex Gaussian distributions with different variances. As a result, its squared norm $\norm{\y - \H\x}^2$ is a random variable defined by the sum of the squares of random variables following Gaussian distributions. To investigate the asymptotic properties of \eqref{eq:p_y_AH1} with respect to SNR, we consider the following random variable $Y$ as a simplified model of \eqref{eq:p_y_AH_numerical} expressed by
\begin{equation}
    Y = -\log_2\qty(\sum_{i=1}^N\exp\qty(-X_i^2)),
\end{equation}
where $X_i$ is a random variable following a Gaussian distribution $\mathcal{N}(0, \sigma_x^2)$. Now, we introduce a new random variable $X_{\mathrm{min}}^2 = \min\qty{X_1^2,\cdots, X_N^2}$. Its expected value $\sigma_{X_{\mathrm{min}}^2}$ is given by
\begin{align}
    \label{sigma_Xmin}
    \sigma_{X_{\mathrm{min}}^2} = \int_{0}^\infty x^2 \frac{d}{dx}\qty{1 - \qty(1 - \Phi\qty(\frac{x}{\sigma_x}))^N}\ dx = \sigma_x^2 g(N)
\end{align}
and
$g(N) = \int_{0}^\infty t^2 \frac{d}{dt}\qty{1 - \qty(1 - \Phi\qty(t))^N}\ dt$,
where $\Phi(\cdot)$ is the cumulative distribution function of a half-Gaussian distribution with unit variance. From \eqref{sigma_Xmin}, $\sigma_{X_{\mathrm{min}}^2}$ increases linearly with SNR. Here, the following inequality holds:
\begin{align}
    Y &= \log_2(e)X_{\mathrm{min}}^2 - \log_2\qty(\sum_{i=1}^N\exp\qty(-X_i^2 + X_{\mathrm{min}}^2)) \nonumber\\
    &\geq \frac{X_{\mathrm{min}}^2}{\log(2)} -\log_2(N).
\end{align}
Since $\log_2\qty(\sum_{i=1}^N\exp\qty(-X_i^2 + X_{\mathrm{min}}^2)) \geq 0$, we have
\begin{equation}
    \frac{X_{\mathrm{min}}^2}{\log(2)} -\log_2(N) \leq Y \leq \frac{X_{\mathrm{min}}^2}{\log(2)}.
\end{equation}
If the sample size $N$ is constant, then $Y \sim X_{\mathrm{min}}^2/\log(2)$ asymptotically holds because of $\sigma_x \gg 0 \Rightarrow X_{\mathrm{min}}^2 \gg \log_2(N)$. Also, for the expected value of $Y$, the following approximation holds:
\begin{equation}
    \sigma_{Y} \sim \frac{\sigma_{X_{\mathrm{min}}^2}}{\log(2)} = \sigma_x^2 \cdot \frac{g(N)}{\log(2)}.
\end{equation}
Thus, asymptotically, $\sigma_Y$ increases linearly with SNR. In other words, $\sigma_Y$ increases exponentially with SNR in units of decibels.

The above analysis explains the phenomenon observed in Fig.~\ref{fig:AMI_error}. The expected value calculation in log-sum-exp, as in \eqref{eq:p_y_AH1}, requires a sufficient sample size because the error increases with SNR.
That is, at high SNRs, it is a challenging task to calculate accurate AMI of GQSM in terms of computational complexity.

\subsection{Closed-Form Expression of \eqref{eq:p_y_AH1}} \label{subsec:closed-form}
The calculation error in \eqref{eq:p_y_AH1} induced by the Monte Carlo integration increases exponentially with SNR.
Here, a closed-form expression of \eqref{eq:p_y_AH1} can be derived, which eliminates the calculation error and provides a more accurate value of AMI.
Assuming that each element of the input symbol independently follows a complex Gaussian distribution $\mathcal{CN}(0, \sigma_s^2/2K)$, \eqref{eq:p_y_AH1} can be expressed as
\begin{equation}
    \label{eq:p_y_AH1_gaussian}
    p(\y | \A, \H) = \frac{\int_{\mathbb{R}^{2K}} \exp\qty(-\frac{\norm{s}^2}{\sigma_{s}^{2}/K} - \frac{\norm{\y - \H\x}^2}{\sigma_n^2})\ d\s}{\qty(\pi \sigma_{s}^{2}/K)^K\qty(\pi \sigma_n^2)^{M_r}}.
\end{equation}
\begin{figure*}[tb]
    \begin{align}
    \label{eq:p_y_AH1_gaussian_ex}
    p(\y | \A, \H)
    &= \frac{1}{\pi^3\sigma_s^2\sigma_n^4} \int_{-\infty}^\infty \int_{-\infty}^\infty \exp\qty(-\frac{s_\R^2 + s_\I^2}{\sigma_{s}^{2}} - \frac{\norm{\y - \H\qty(\A_\R\s_\R + \j\A_\I\s_\I)}^2}{\sigma_n^2})\ ds_\R ds_\I \nonumber\\
    &= \frac{1}{\pi^3\sigma_s^2\sigma_n^4}  \int_{-\infty}^\infty \int_{-\infty}^\infty \exp\qty(-\frac{p_1 s_\R^2}{\sigma_s^2\sigma_n^2} + \frac{2s_\R}{\sigma_n^2}\qty(\alpha s_\I + \beta) - \frac{1}{\sigma_n^2}\qty(\norm{\y}^2 + 2\gamma s_\I + \frac{p_2}{\sigma_s^2}s_\I^2))\ ds_\R ds_\I
\end{align}
\end{figure*}
In general, \eqref{eq:p_y_AH1_gaussian} can be expressed in closed form by repeating the Gaussian integral $2K$ times. By substituting the closed-form expression of \eqref{eq:p_y_AH1_gaussian} into \eqref{eq:I_continuous}, the AMI of GQSM for continuous-input channels can be obtained with minimal error.
Although the closed form of the expected value $\E_\s[\cdot]$ can be obtained in \eqref{eq:I_continuous}, it is still necessary to calculate $\E_{\A, \H, \y}[\cdot]$ by the Monte Carlo method, which induces no significant calculation errors.
In repeating the Gaussian integral $2K$ times, we used symbolic computation with the computer algebra system GiNaC~\cite{vollinga2006ginac} and the linear algebra library Eigen~\cite{eigenweb}.

As an example, in the case of $(N_t, N_r, K, \mathbf{A}_\R, \mathbf{A}_\I) = (2, 2, 1, [0\ 1]^\mathrm{T}, [1\ 0]^\mathrm{T})$, \eqref{eq:p_y_AH1_gaussian} is given by \eqref{eq:p_y_AH1_gaussian_ex}, where $\mathbf{h}_i$ is the $i$-th column vector of $\H$, and $h_{\R}^{(i, k)}$ and $h_{\I}^{(i, k)}$ are the real and imaginary parts of the $(i, k)$ element of $\H$.
The real and imaginary parts of the input symbol $s$ are denoted by $s_\R$ and $s_\I$, while the real and imaginary parts of the $i$-th element of the received signal $\mathbf{y}$ are denoted by $y_{\R}^{(i)}$ and $y_{\I}^{(i)}$, respectively.
The constant parameters in \eqref{eq:p_y_AH1_gaussian_ex} are defined as
\begin{align}
    p_1 &= \sigma_s^2\norm{\mathbf{h}_1}^2 + \sigma_n^2, ~ ~ p_2 = \sigma_s^2\norm{\mathbf{h}_2}^2 + \sigma_n^2,\\
    \alpha &= \textstyle{\sum_{i=1}^2 \qty(h_{\R}^{(i, 1)} h_{\I}^{(i, 2)} - h_{\I}^{(i, 1)} h_{\R}^{(i, 2)})},\\
    \beta &= \textstyle{\sum_{i=1}^2\qty(h_{\R}^{(i, 1)}y_{\R}^{(i)} + h_{\I}^{(i, 1)}y_{\I}^{(i)})}, ~ \mathrm{and} \\
    \gamma &= \textstyle{\sum_{i=1}^2 \qty(h_{\I}^{(i, 2)} y_{\R}^{(i)} - h_{\R}^{(i, 2)} y_{\I}^{(i)})}.
\end{align}
Integrating \eqref{eq:p_y_AH1_gaussian_ex} with respect to $s_\R$ gives
\begin{equation}
    \label{eq:p_y_AH1_gaussian_ex2}
    p(\y | \A, \H)
    = \frac{1}{\pi^{\frac{5}{2}}\sigma_s\sigma_n^{3}\sqrt{p_1}} \int_{-\infty}^\infty \exp\qty(Z_1)\ ds_\I,
\end{equation}
where
\begin{equation}
\begin{split}
     Z_1 &= -\frac{s_\I^2}{\sigma_n^2}\qty(\frac{p_2}{\sigma_s^2}  - \frac{\sigma_s^2\alpha^2}{p_1}) + \frac{2s_\I}{\sigma_n^2}\qty(\frac{\sigma_s^2\alpha\beta}{p_1} - \gamma)\\
     &\quad + \frac{1}{\sigma_n^2}\qty(\frac{\sigma_s^2\beta^2}{p_1} - \norm{\mathbf{y}}^2).
\end{split}
\end{equation}
Finally, integrating \eqref{eq:p_y_AH1_gaussian_ex2} with respect to $s_\I$ gives the closed-form expression of \eqref{eq:p_y_AH1_gaussian} as
\begin{equation}
    p(\y | \A, \H)
    = \frac{\exp\qty(Z_2)}{\pi^2\sigma_n^2\sqrt{p_1 p_2 - \sigma_s^4\alpha^2}},
\end{equation}
where
\begin{equation}
 Z_2 = \frac{1}{\sigma_n^2}\qty(\frac{p_2}{\sigma_s^2} - \frac{\sigma_s^2\alpha^2}{p_1})\qty(\frac{\sigma_s^2\alpha\beta}{p_1} + \gamma) + \frac{\sigma_s^2\beta^2}{p_1\sigma_n^2} + \frac{\norm{\y}^2}{\sigma_n^4}.
\end{equation}

\section{Numerical Results}
\begin{figure}[tb]
    \centering
    \includegraphics[keepaspectratio,scale=0.69]{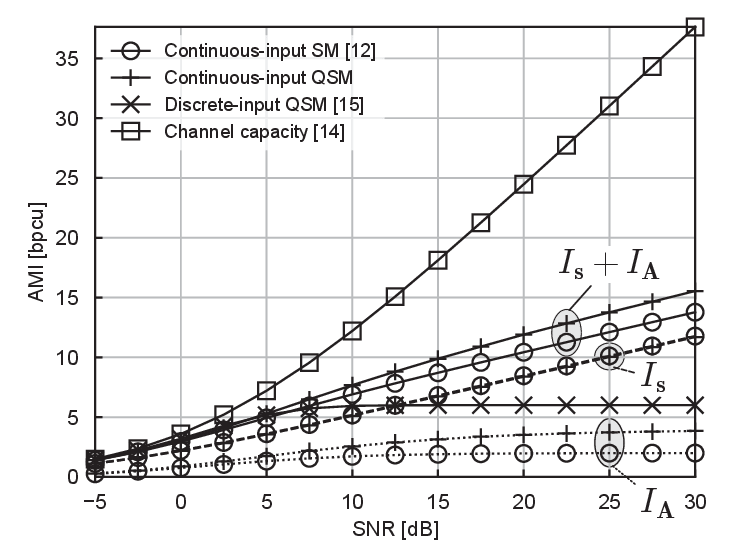}
    \caption{AMI of SM and QSM, where $(N_t, N_r, K, Q) = (4, 4, 1, 16)$. QPSK signaling was considered for discrete-input QSM.}
    \label{fig:SNR_AMI_QSM}
\end{figure}
\begin{figure}[tb]
    \centering
    \includegraphics[keepaspectratio,scale=0.69]{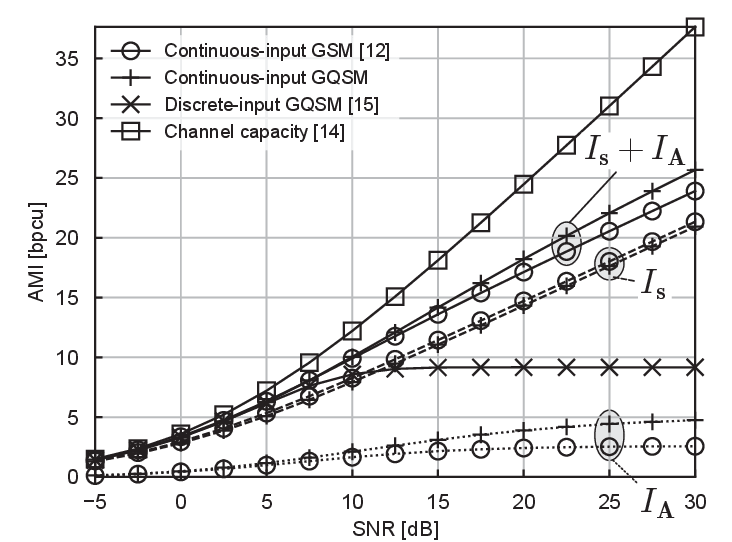}
    \caption{AMI of GSM and GQSM, where $(N_t, N_r, K, Q) = (4, 4, 2, 36)$. QPSK signaling was considered for discrete-input GQSM.}
    \label{fig:SNR_AMI_GQSM}
\end{figure}
\begin{figure}[tb]
    \centering
    \includegraphics[keepaspectratio,scale=0.69]{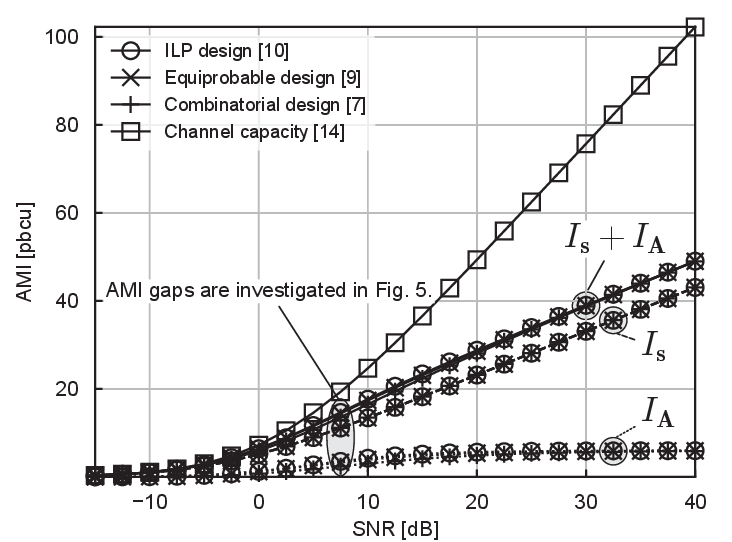}
    \caption{AMI of GQSM for different activation patterns~\cite{basar2013orthogonal,wen2016equiprobable,ishikawa2019imtoolkit} where $(N_t, N_r, K, Q) = (8, 8, 3, 64)$.}
    \label{fig:SNR_AMI_GQSM_AP}
\end{figure}
\begin{figure}[tb]
    \centering
    \includegraphics[keepaspectratio,scale=0.69]{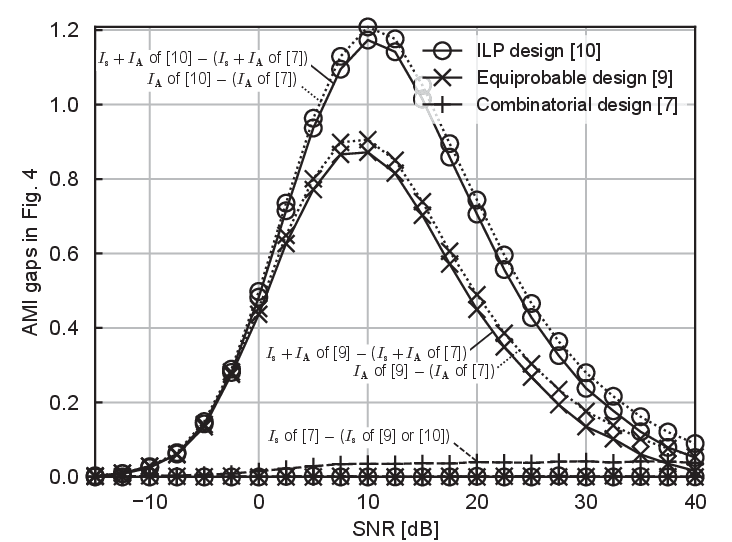}
    \caption{AMI gaps of GQSM for different activation patterns~\cite{basar2013orthogonal,wen2016equiprobable,ishikawa2019imtoolkit} where $(N_t, N_r, K, Q) = (8, 8, 3, 64)$.}
    \label{fig:SNR_AMI_GQSM_AP_differences}
\end{figure}
In this section, we compare the AMI of QSM and GQSM with those of related SM schemes.\footnote{The AMI of QSM derived in \cite{younis2017quadrature} is omitted here since it is equal to the MIMO channel capacity~\cite{telatar1999capacity} when $N_t \rightarrow \infty$.} All the results were obtained through Monte Carlo simulations over $10^6$ independent channel realizations. Each element of the input symbols was assumed to independently follow a complex Gaussian distribution.

First, Fig.~\ref{fig:SNR_AMI_QSM} shows the AMI of QSM and SM, where $(N_t, N_r, K, Q) = (4, 4, 1, 16)$. The dashed and dotted lines in the figure correspond to $I_\s$ and $I_\A$ in \eqref{eq:I_continuous}, respectively. The figure shows that the difference in AMI between QSM and SM can be mainly attributed to $I_\A$, and there was little difference in $I_\s$. Since $I_\A$ is based on a finite number of APs, it converged to $\log_2 Q$ at high SNRs.

Next, Fig.~\ref{fig:SNR_AMI_GQSM} shows the AMI of GSM and GQSM, where $(N_t, N_r, K, Q) = (4, 4, 2, 36)$. Note that $I_\s$ of GSM is equal to the ergodic capacity of $K \times N_r$ MIMO. Although the results were basically the same as those shown in Fig.~\ref{fig:SNR_AMI_QSM}, $I_\s$ of GQSM was slightly smaller than that of GSM. This can be attributed to the fact that GQSM is more susceptible to channel fading than GSM when the APs of the real and imaginary parts do not match, i.e., $\A_\R \neq \A_\I$. Specifically, GSM is affected by $K$ channel elements, while GQSM can be affected by up to $2K$ channel elements. Interestingly, in the SNR range below $7.5$ dB, the decrease in $I_\s$ exceeded the increase in $I_\A$, resulting in the AMI of GQSM being lower than that of GSM.\footnote{Since $Q = \binom{N_t}{K}^2$ holds in Figs.~2 and 3, the number of APs is equal to the cardinality of all candidates, and there is no room for designing APs.}

Finally, Fig.~\ref{fig:SNR_AMI_GQSM_AP} shows the AMI of GQSM for different AP designs~\cite{basar2013orthogonal,wen2016equiprobable,ishikawa2019imtoolkit}, where $(N_t, N_r, K, Q) = (8, 8, 3, 64)$. For simplicity, the same APs were used for both the real and imaginary parts of the codewords, i.e., $\mathbb{A}_\R = \mathbb{A}_\I$.
As shown, the differences in AMI appeared to be small. To further analyze these differences, in Fig.~\ref{fig:SNR_AMI_GQSM_AP_differences},
we focus on the differences in AMI between the three AP designs, which shows that the differences in AMI depended on the SNR and were maximized at medium SNRs. The ILP~\cite{ishikawa2019imtoolkit} and equiprobable~\cite{wen2016equiprobable} designs maximized the equiprobability of the active transmit antennas and also increased the probability of being affected by channel fading, leading to a slight decrease in $I_\s$. However, the decrease in $I_\s$ was negligible compared with the increase in $I_\A$, resulting in overall improvements in AMI compared with the combinatorial design~\cite{basar2013orthogonal}.

\section{Conclusions\label{sec:conc}}
In this letter, we derived the AMI of GQSM for continuous-input channels, which clarified a significant difference in AMI between GQSM and GSM at high SNRs. Additionally, the impact of AP designs on AMI was maximized at medium SNRs, and the maximum AMI was achieved by the ILP design. The analyses given in this letter are applicable to the schemes subsumed by GQSM, such as QSM and OFDM-I/Q-IM.

\footnotesize{
    \bibliographystyle{IEEEtranURLandMonthDiactivated}
	\bibliography{main}
}

\end{document}